\pdfoutput=1

\documentclass[sigconf]{acmart}

\usepackage{booktabs} 
\usepackage{tikz}
\usetikzlibrary{shapes.geometric, arrows, fit}
\usepackage{subcaption}
\usepackage{caption}
\usepackage[export]{adjustbox}

\setcopyright{rightsretained}

\begin{document}
\title{cGAN-based Manga Colorization Using a Single Training Image}

 \author{Paulina Hensman}
 \authornote{Exchange from Royal Institute of Technology, Sweden}
 \orcid{1234-5678-9012}
 \affiliation{%
   \department{Dept. Information and\\ Communication Eng.}
   \institution{University of Tokyo}
 }
 \email{phensman@kth.se}

 \author{Kiyoharu Aizawa}
 \affiliation{%
   \department{Dept. Information and\\ Communication Eng.}
   \institution{University of Tokyo}
 }
 \email{aizawa@hal.t.u-tokyo.ac.jp}


\begin{abstract} 
The Japanese comic format known as Manga is popular all over the world. It is traditionally produced in black and white, and colorization is time consuming and costly. Automatic colorization methods generally rely on greyscale values, which are not present in manga. Furthermore, due to copyright protection, colorized manga available for training is scarce.

We propose a manga colorization method based on conditional Generative Adversarial Networks (cGAN). Unlike previous cGAN approaches that use many hundreds or thousands of training images, our method requires only a single colorized reference image for training, avoiding the need of a large dataset. 

Colorizing manga using cGANs can produce blurry results with artifacts, and the resolution is limited. We therefore also propose a method of segmentation and color-correction to mitigate these issues. The final results are sharp, clear, and in high resolution, and stay true to the character's original color scheme.

\end{abstract}

%
%




\maketitle

\section{Introduction}

Manga is the main comic style used in Japan, and it is enjoyed by all ages across the world. It is produced at a rapid pace, and is traditionally drawn in black and white. Instead of classic shading, screentones are used to show light and texture.

A colorized manga is often more visually appealing, and with digital distribution methods color printing costs are no longer an issue. That being said, the actual colorization of the image still takes time and requires a skilled artist. Any way to automate this process could potentially be very helpful in production of colorized manga.

While a fully automatic method for manga colorization would be optimal, some user input will always be required for a correct output. Certain parts of a manga image such as skin color and background elements could possibly be correctly estimated and colorized by a good method. However, hair color and other attributes that make a character unique can not possibly be inferred from a monochrome image. We therefore propose a method using a colorized image as reference to colorize other images of the same character. 

\begin{figure}
\includegraphics[width=\linewidth]{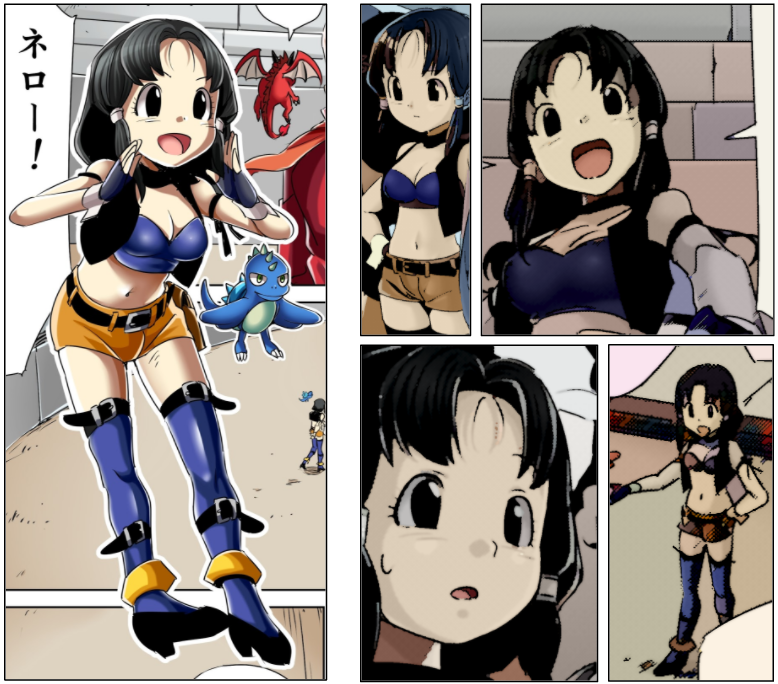}
\begin{subfigure}[t]{.4\linewidth}
\caption{Training image}
\end{subfigure}\hfill
\begin{subfigure}[t]{.55\linewidth}
\caption{Our results}
\end{subfigure}
\caption{On the left, a colorized training image. On the right, images colorized by our method. \copyright Whomor \label{potential}}
\end{figure}

The same characters tend to recur many times in a manga, especially if the manga goes on for several volumes. Additionally, uniforms such as school uniforms, team uniforms, and military uniforms are common. Requiring only a single reference image for each character or each uniform could thus save a lot of time when colorizing a manga. An example of a recurring character is shown in Fig.~\ref{potential}.

Comics are protected by copyright as works of art, making it difficult to gather large datasets of manga to use for training. The largest available dataset is Manga109\cite{matsui2016sketch}\cite{manga109}, which unfortunately contains only monochrome images. This is a constraint for manga colorization methods, as methods that rely heavily on large amounts of training data can not be used. 

We propose a solution for the task of manga colorization using conditional Generative Adversarial Networks (cGANs). Unlike previous cGAN literature, we demonstrate convincing results using a single training image of the character to be colorized. This eliminates the need for a large dataset of colorized manga unprotected by copyright, while also using the color palette specific to the character during colorization. Our method could potentially be used by manga artists or, with permission, by their fans to colorize images, creating more colorized works for fans of the genre to enjoy.

The contributions of our paper are summarized below.

\begin{itemize}
    \item We show that a single colorized image is enough to train a cGAN for the task of manga colorization for a specific character.
    \item We propose a chain of post-processing methods to refine noisily colorized manga images.
    \item Put together, we introduce a full system for manga colorization. It is the first method for manga colorization that does not require user interaction.
    \item We use two different datasets, with and without screentones, to show the effectiveness of our method in both cases.
\end{itemize}

\section{Related work}
\subsection{Color transfer methods for natural images}
There are many existing methods for color transfer, i.e. using the colors from one image to colorize another\cite{reinhard2001color}\cite{Hristova2015Style}. The common thread they share is that they make use of preexisting information in the image to be colored, such as color information for a color transfer or luminance in a greyscale image to be colored. Manga is however in its pure form a binary image of only black and white. While shading exists, it comes from screentones which is again just black patterns on white. Thus, the information those methods rely on is not available.

\subsection{Style transfer methods}
In a similar vein to color transfer, style transfer attempts to transfer the style from a reference image onto a new image. Regarding monochrome manga and colorized manga as two different styles, this method could allow us to transform a monochrome manga image into a colorized one. Notable work in this field has been done by Hertzmann et al.\cite{hertzmann2001image} and Chuan Li et al.\cite{imgSynth}. 

\subsection{Conditional Generative Adversarial Networks}
The recently introduced Generative Adversarial Networks (GAN)\cite{goodfellow2014generative} by Goodfellow et al. have shown promising results in generating vividly colorful images\cite{radford2015unsupervised}\cite{reed2016generative}\cite{dosovitskiy2016generating}. A notable recent example is the the conditional Generative Adversarial Network (cGAN) presented by Isola et al.\cite{isola2016image}, which generates new images by conditioning on a target image. This method successfully learned a mapping between edge lines and image for objects such as shoes and handbags. The edge lines are similar to our monochrome manga images, indicating that similar results could be achieved for our task using this method. Up until now, large amounts of training data have been used to train this type of network for generalizing. In the original paper, the smallest training set mentioned consisted of 400 images. We will show that a single colorized reference image is enough to train such a network to colorize a specific character.

\subsection{Manga colorization}

There have only been a few works on manga colorization. As manga images consist of only black lines and screentones on white background, they are different in nature from natural images, making them more difficult to work with. 

Qu et al.\cite{qu2006manga} proposed a method to colorize manga using hint color scribbles manually added by the user. The method required a lot of user interaction for each image to add the hint colors.  

A method for manga colorization based on a reference image was proposed by Sato et al.\cite{sato2014reference}. Their approach used graph matching to match segments of the reference image to the target image. Their method was limited to simple images.

\tikzstyle{startstop} = [rectangle, rounded corners, minimum width=2cm, minimum height=0.6cm,text centered, draw=black, fill=red!30]
\tikzstyle{io} = [rectangle, rounded corners, minimum width=1cm, minimum height=0.6cm, text centered, text width=2cm, draw=black, fill=blue!30]
\tikzstyle{process} = [rectangle, rounded corners, minimum width=3cm, minimum height=0.6cm, text centered, draw=black]
\tikzstyle{training} = [rectangle, minimum width=2.5cm, minimum height=0.6cm, text centered, draw=black, fill=gray!50]
\tikzstyle{arrow} = [thick,->, shorten >=1pt] 

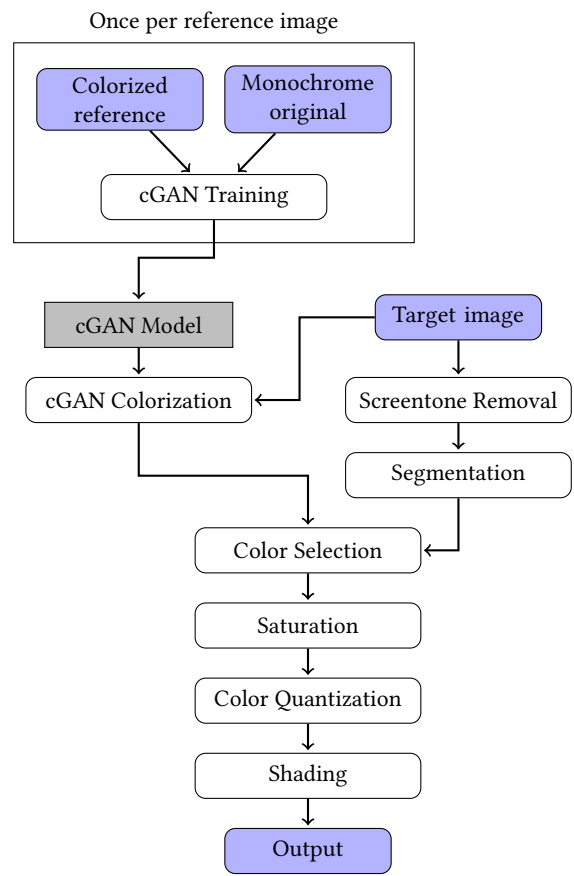
\begin{figure}
\begin{tikzpicture}[node distance=1cm]

\node (in1) [io] {Colorized reference};
\node (in3) [io, xshift=2.5cm] {Monochrome original};
\node (train) [process, below of=in3, xshift=-1.25cm, yshift=-0.3cm] {cGAN Training};
\node (model) [training, below of=train, yshift=-0.7cm, xshift=-1cm] {cGAN Model};

\node (in2) [io, xshift=4.5cm, yshift=-2.9cm] {Target image};
\node (screenr) [process, below of=in2, yshift=-0.1cm] {Screentone Removal};
\node (seg) [process, below of=screenr] {Segmentation};

\node (sty) [process, below of=model] {cGAN Colorization};
\node (col) [process, below of=seg, xshift=-2cm] {Color Selection};
\node (sat) [process, below of=col] {Saturation};
\node (quant) [process, below of=sat] {Color Quantization};
\node (screen) [process, below of=quant] {Shading};

\node (stop) [io, below of=screen] {Output};

\draw [arrow] (in1) -- (train);
\draw [arrow] (in3) -- (train);

\draw [arrow] (train.south) -- ++(0,-0.5) -- ++(-1,0) --  (model.north);

\draw [arrow] (model) -- (sty);
\draw [arrow] (in2.west) -- ++(-1,0) -- ++(0,-1.1) --  (sty.east);
\draw [arrow] (in2) -- (screenr);
\draw [arrow] (screenr) -- (seg);

\draw [arrow] (sty.south) -- ++(0,-0.7) -- ++(2.25,0) --  (col.north);
\draw [arrow] (seg) |- (col);
\draw [arrow] (col) -- (sat);
\draw [arrow] (sat) -- (quant);
\draw [arrow] (quant) -- (screen);
\draw [arrow] (screen) -- (stop);

\node[draw,inner sep=0.3cm,label=above:Once per reference image,fit=(in1) (in3) (train)] {};

\end{tikzpicture}
\caption{A diagram of the proposed method. \label{diagram}}
\end{figure}

\section{Method}
Our method consists of two main parts, cGAN colorization and post-processing. A diagram of the method is shown in Fig.~\ref{diagram}, and a sample output image from each step is shown in Fig.~\ref{eachstep}. The required training data is a single colorized image and its corresponding monochrome original. Input data is a monochrome target image.

Some parameters have different optimal settings depending on the image, and are thus available for adjustment by the user. An overview of the tunable parameters is shown in Table~\ref{tunable}. 

\subsection{Dataset}
The data used comes from two sources. 

\begin{figure}
\includegraphics[width=.50\linewidth]{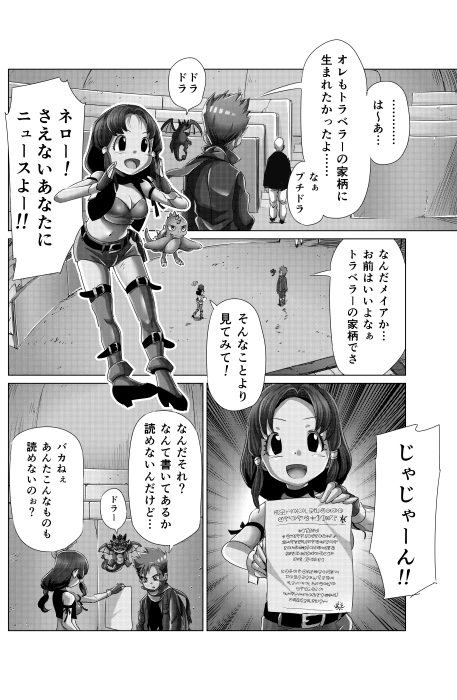}\hfill
\includegraphics[width=.50\linewidth]{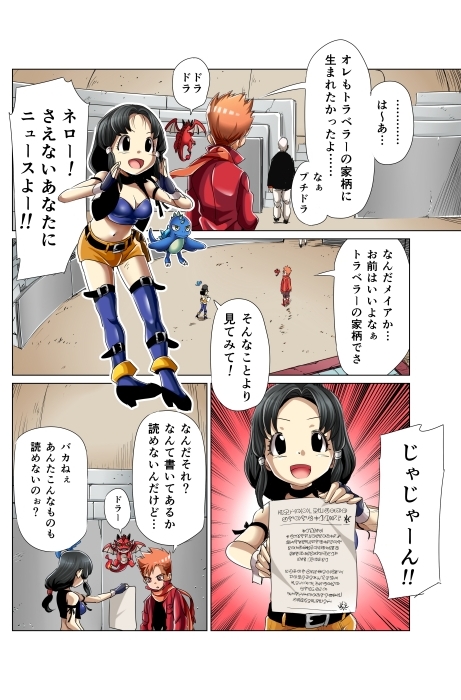}
\caption{A sample page from the Monster Beat dataset, monochrome and colorized. \label{whomorex}}
\end{figure}

We have used 8 pages of the manga "Monster Beat", where each page has both a colorized and a monochrome version. See Fig.~\ref{whomorex} for a sample page. The manga contains 2 prominent characters, one male and one female. The male character appears 18 times in the set, while the female appears 11 times. There is also an older male character appearing 6 times. Appearances include face, half-body, and full-body. 

\begin{figure}
\includegraphics[width=.50\linewidth]{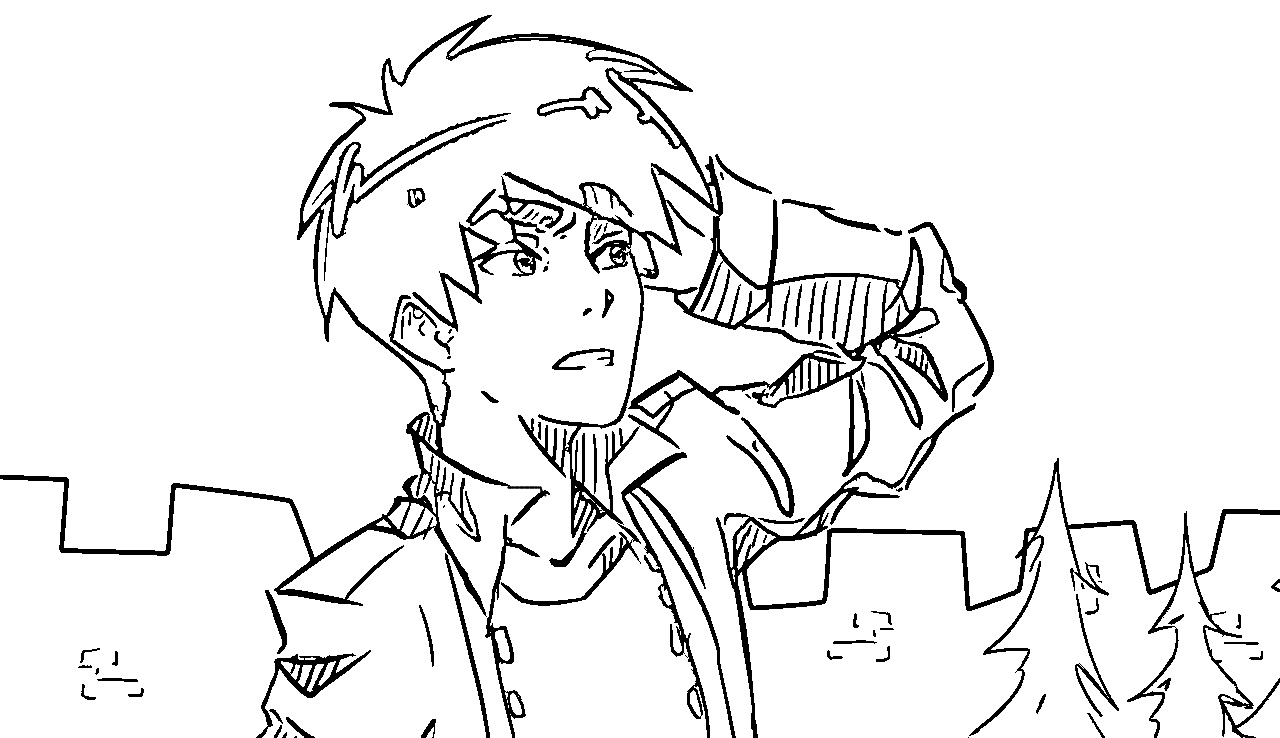}\hfill
\includegraphics[width=.50\linewidth]{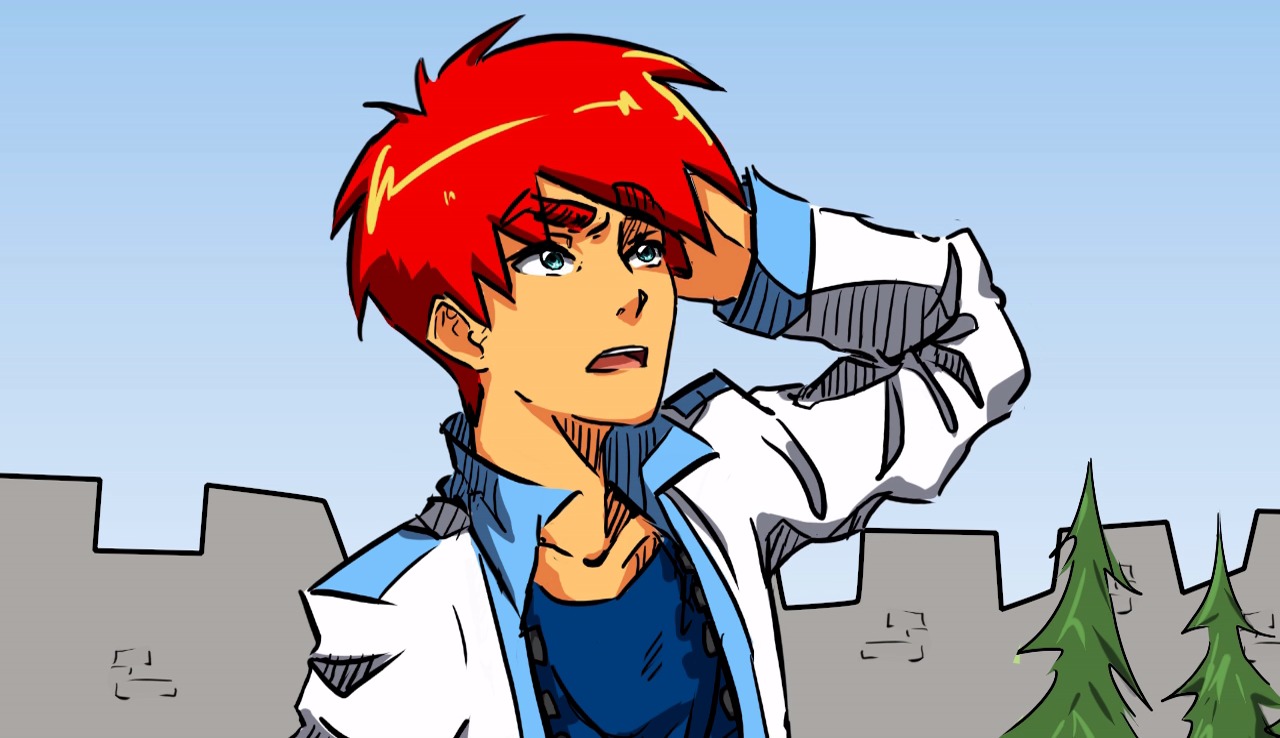}
\caption{A sample frame from the Morevna dataset\cite{morevna}, with a matching generated line drawing. \label{morevnaex}}
\end{figure}

We have also used frames from project Morevna\cite{morevna}, an animated cartoon drawn in manga style. Morevna consists of two videos summing up to a total of 15 minutes, featuring the same two main characters drawn in slightly different styles. At two frames per second, removing all black frames, we created a dataset of 1600 images from the videos. Examples are shown in Fig.~\ref{morevnaex}. As all images were already colorized, we used the screen tone removal method described in section~\ref{sec:screentone} to create monochrome line drawings from the images.

\subsection{Screentone removal}\label{sec:screentone}
The screentones used for shading and texture in manga can be helpful or unhelpful when colorizing. In the cGAN colorization step, the screentones work as hints, as areas with the same type of screentone tend to have the same color. During our segmentation method however, the gaps in the screentones form extra segments, dividing up the true segments. We thus remove the screentones before performing the segmentation. 

To extract screentones, we use the method proposed by Ito et al.\cite{ito2015separation}. However, if a screentone-free version of the target image is available, which might be the case if the user is the original artist, supplying this version as well gives cleaner lines in the final output.

\begin{figure}
\includegraphics[width=.20\linewidth]{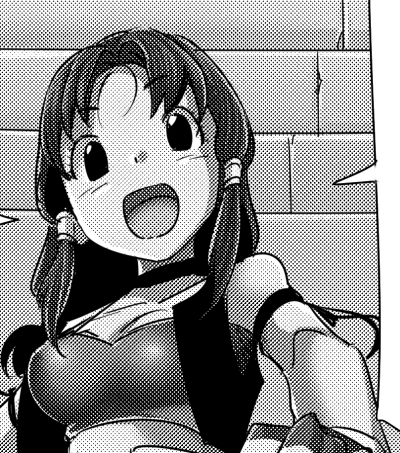}\hfill
\includegraphics[width=.20\linewidth]{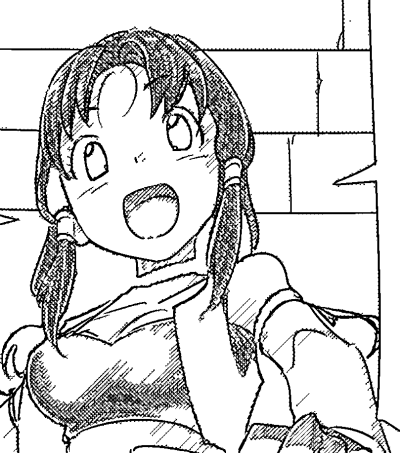}\hfill
\includegraphics[width=.20\linewidth]{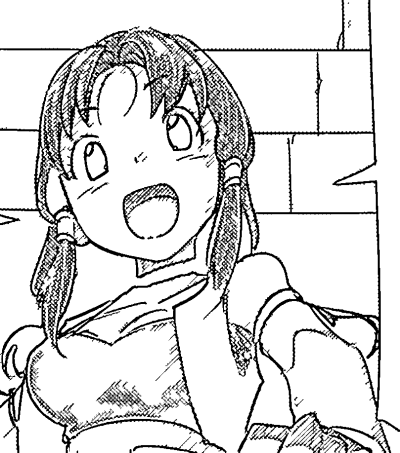}\hfill
\includegraphics[width=.20\linewidth]{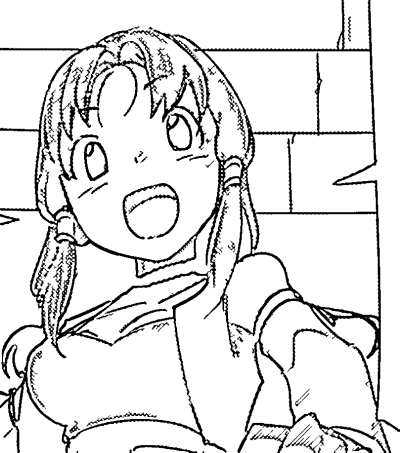}\hfill
\includegraphics[width=.20\linewidth]{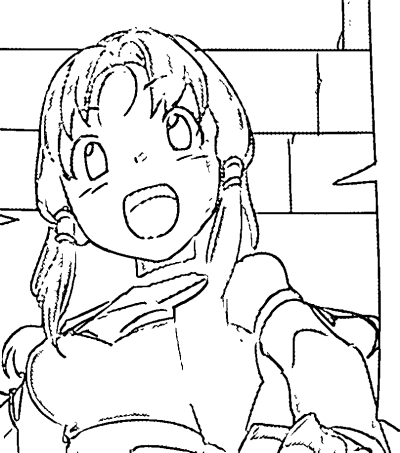}
\caption{An original monochome image on the left, followed by the results of tone removal after no Gaussian blur, and after Gaussian blur with a kernel radius of 1, 2 and 3 respectively. \label{gauss}}
\end{figure}

When working with images in high resolution, screentones will sometimes not be completely removed by Ito et al.'s method. This is presumably due to the sharpness of the screentones causing them to be confused with edge lines. Applying a Gaussian blur to the images before removing the screentones produced better results to a certain extent. As seen in Fig.~\ref{gauss}, Gaussian blur with a kernel radius of 1 or 2 gave improved results, while at a kernel radius of 3 the outlines started disappearing, reducing the quality of the result. We can see that there is an optimal value, and our experiments show that it depends on the input image. We therefore make this setting available for adjustment by the user.

\subsection{cGAN colorization}
\begin{figure*}
\begin{subfigure}[t]{.10\linewidth}
\vspace{1.7cm}

Face

\vspace{2.3cm}

Chest

\vspace{2.3cm}

Half body

\vspace{2.3cm}

Full body

\vspace{2.3cm}

All
\end{subfigure}
\begin{subfigure}[t]{.20\linewidth}
\includegraphics[width=.72\linewidth]{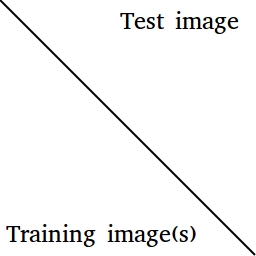}\\\\
\includegraphics[width=.72\linewidth]{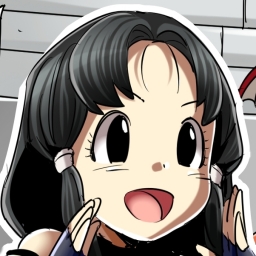}

\vspace{0.097cm}

\includegraphics[width=.72\linewidth]{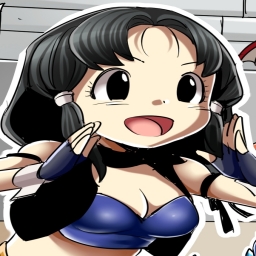}

\vspace{0.097cm}

\includegraphics[width=.72\linewidth]{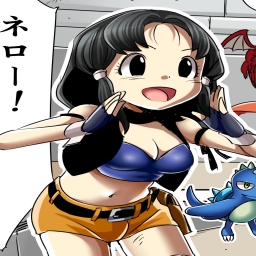}

\vspace{0.097cm}

\includegraphics[width=.72\linewidth]{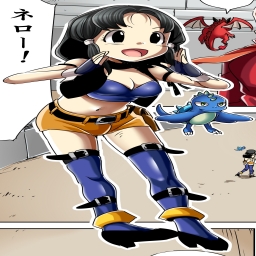}

\vspace{0.097cm}

\includegraphics[width=.72\linewidth]{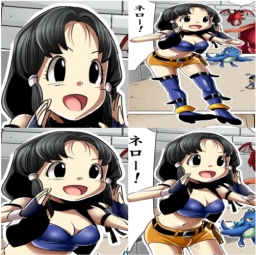}\\
\end{subfigure}
\begin{subfigure}[t]{.60\linewidth}
\includegraphics[width=.24\linewidth]{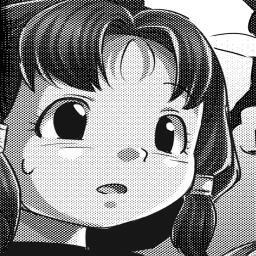}\hfill
\includegraphics[width=.24\linewidth]{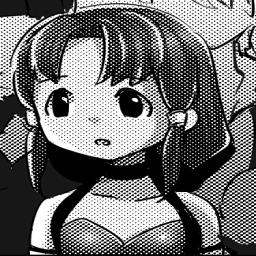}\hfill
\includegraphics[width=.24\linewidth]{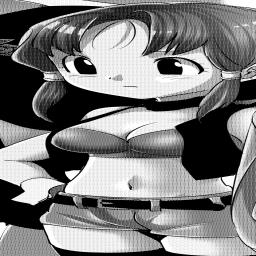}\hfill
\includegraphics[width=.24\linewidth]{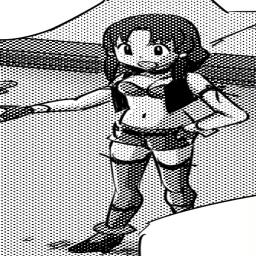}\hfill\\\\
\includegraphics[width=.23\linewidth, cfbox=green 2pt 0pt]{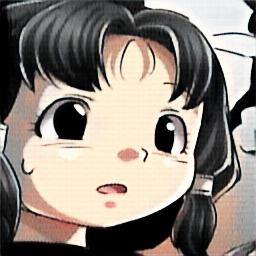}\hfill
\includegraphics[width=.24\linewidth]{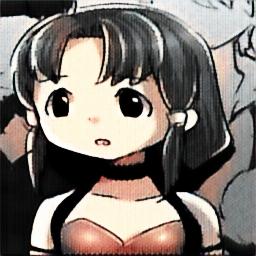}\hfill
\includegraphics[width=.24\linewidth]{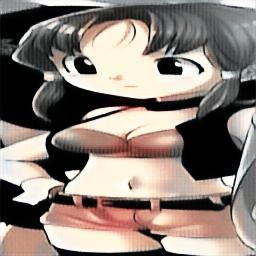}\hfill
\includegraphics[width=.24\linewidth]{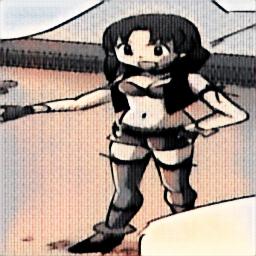}\hfill\\
\includegraphics[width=.24\linewidth]{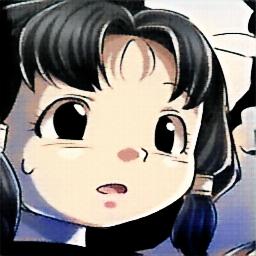}\hfill
\includegraphics[width=.23\linewidth,cfbox=green 2pt 0pt]{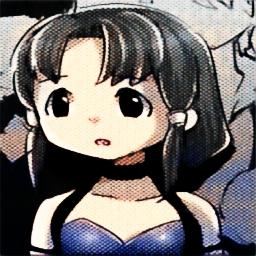}\hfill
\includegraphics[width=.24\linewidth]{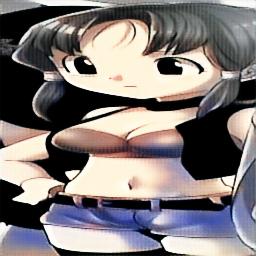}\hfill
\includegraphics[width=.24\linewidth]{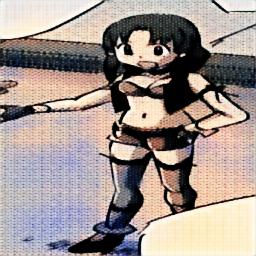}\hfill\\
\includegraphics[width=.24\linewidth]{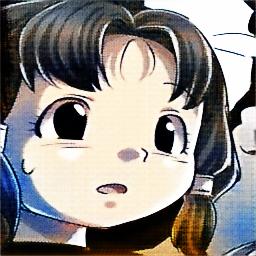}\hfill
\includegraphics[width=.24\linewidth]{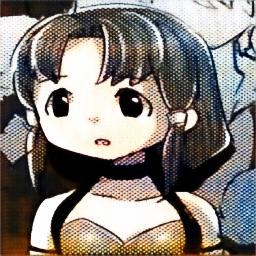}\hfill
\includegraphics[width=.23\linewidth, cfbox=green 2pt 0pt]{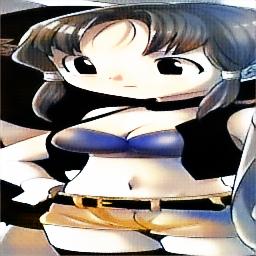}\hfill
\includegraphics[width=.24\linewidth]{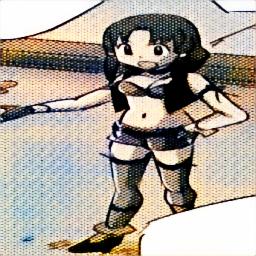}\hfill\\
\includegraphics[width=.24\linewidth]{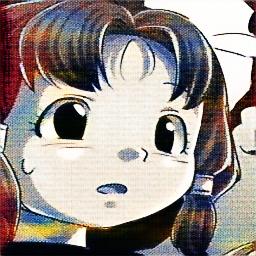}\hfill
\includegraphics[width=.24\linewidth]{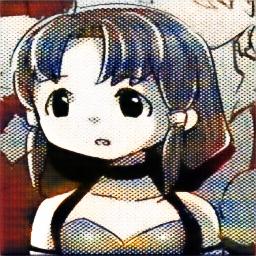}\hfill
\includegraphics[width=.24\linewidth]{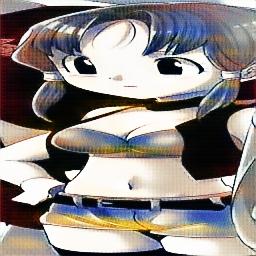}\hfill
\includegraphics[width=.23\linewidth, cfbox=green 2pt 0pt]{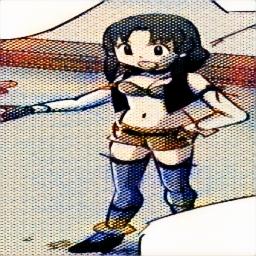}\hfill\\
\includegraphics[width=.24\linewidth]{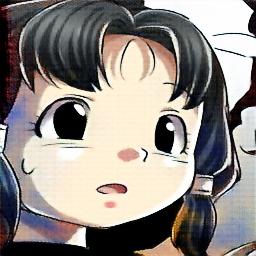}\hfill
\includegraphics[width=.24\linewidth]{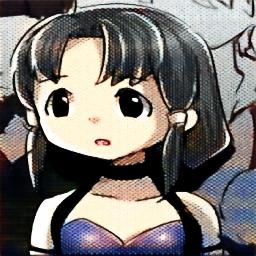}\hfill
\includegraphics[width=.24\linewidth]{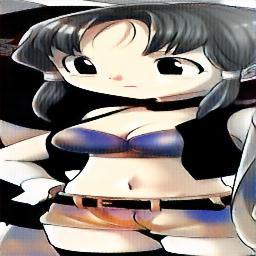}\hfill
\includegraphics[width=.23\linewidth, cfbox=green 2pt 0pt]{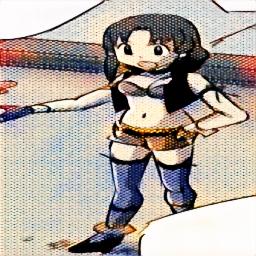}\hfill
\end{subfigure}
\caption{Comparison between combinations of training image and test image for cGAN colorization results.  \label{traindata}}
\end{figure*}

We use Isola et al.'s\cite{isola2016image} method for an initial colorization of the images. The training is done with a single pair of images: a colorized reference image and a corresponding monochrome image. The monochrome image can be generated from the colorized image if no original is available. If the target images have screentones, converting the colorized image to greyscale is sufficient. Otherwise, Ito et al.'s\cite{ito2015separation} screentone removal method can be used to create line drawings as we did for the Morevna dataset. 

We found that the coloring stabilizes after training for less than 100 iterations, while further training sharpens the edges of the output image. The results shown in this paper are based on training with 500 iterations, but results after 100-200 iterations are not significantly worse.  

As we use only a single training image, this method is in its nature inflexible, which means that it is beneficial to use reference images as similar to the image to be colorized as possible. This does not mean several reference images per character are required, however. Starting with a full body image (or less, if the character never appears in full body) of the character, this image can be cropped to show only face, face and upper body, etc. 

Fig.~\ref{traindata} shows results from training with 4 different crops of a reference image on 4 different testing images, plus the results for a network trained with all 4 crops of the image. There is a clear diagonal showing that a crop of only the face is best for colorizing a face, and so on, the exception being the full body image which we considered a tie between the full body network and the combination network. The combination network was second best in all other categories. Backed by these results, we recommend either training several networks on different crops of the reference image, or training a single network on several crops. 

Due to the dataset consisting of only a single image, training is fast, completing 100 iterations in less than 3 minutes on a K80 Tesla GPU. It would not be infeasible to run the training on a CPU.

\subsection{Segmentation} \label{sec:seg}
Manga images for the most part have a natural decomposition into segments, for which each segment has a single color if shading is disregarded. We divide the target image into segments and select a single color for each segment to remove any blur or artifacts created in the cGAN colorization step. For this step, we remove the screentones from the target image, as the spaces formed by the screentones would otherwise form many tiny segments.

\begin{figure*}
\captionsetup{justification=centering}
\begin{subfigure}{.12\linewidth}
    \includegraphics[width=\linewidth]{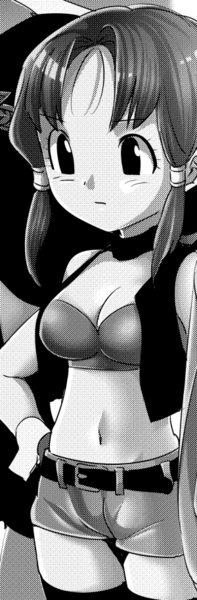}
    \caption{Target\\image}
\end{subfigure}
\begin{subfigure}{.12\linewidth}
    \includegraphics[width=\linewidth]{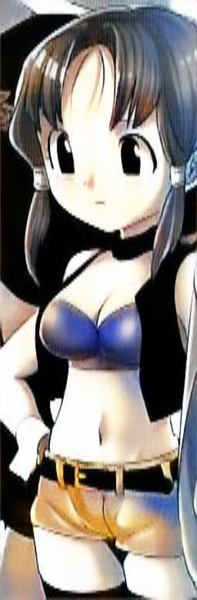}
    \caption{cGAN \\colorization}
\end{subfigure}
\begin{subfigure}{.12\linewidth}
    \includegraphics[width=\linewidth]{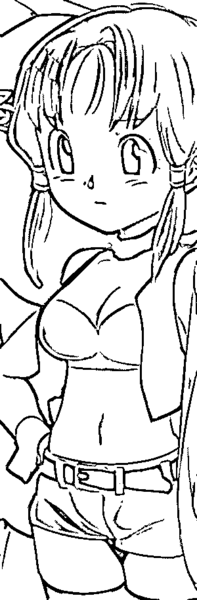}
    \caption{Screentone\\ removal}
\end{subfigure}
\begin{subfigure}{.12\linewidth}
    \includegraphics[width=\linewidth]{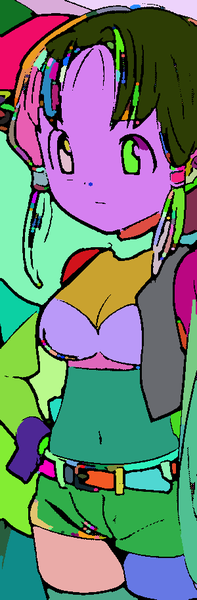}
    \caption{Segmentation\\ \label{segmentation}}
\end{subfigure}
\begin{subfigure}{.12\linewidth}
    \includegraphics[width=\linewidth]{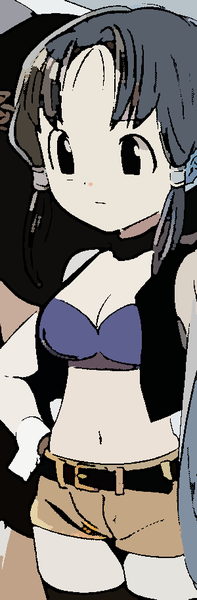}
    \caption{Color \\selection}
\end{subfigure}
\begin{subfigure}{.12\linewidth}
    \includegraphics[width=\linewidth]{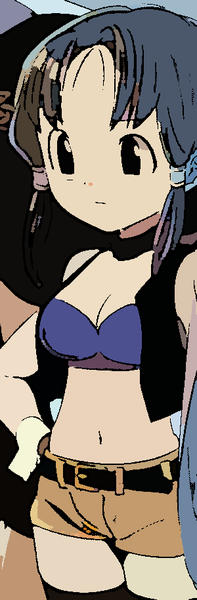}
    \caption{Saturation\\ increase}
\end{subfigure}
\begin{subfigure}{.12\linewidth}
    \includegraphics[width=\linewidth]{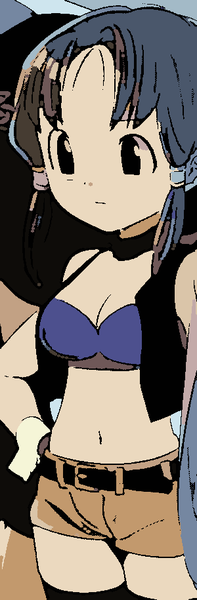}
    \caption{Color\\ quantization \label{quant}}
\end{subfigure}
\begin{subfigure}{.12\linewidth}
    \includegraphics[width=\linewidth]{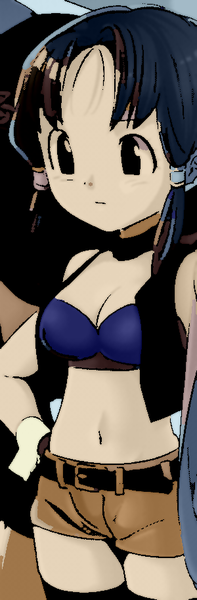}
    \caption{Shading \\addition}
\end{subfigure}


\caption{Output from each step in the method. \label{eachstep}}
\end{figure*}

\begin{figure}
\includegraphics[width=.40\linewidth]{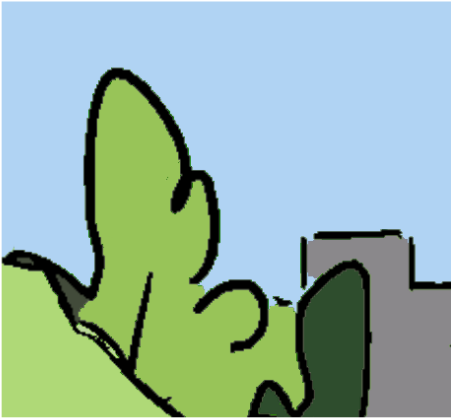}\hfill
\caption{Segments properly separated by the trapped-ball method despite gaps in the edges. \label{gap}}
\end{figure}

We implemented \emph{trapped-ball segmentation} as proposed in \cite{zhang2009vectorizing}, with a few changes. The method works by, metaphorically, placing a ball of a certain diameter in the image. The pixels that the ball can get to without crossing an edge belong to the same segment. This means that any gap between edges narrower than the ball's diameter will be ignored, preventing leaking. See Fig.~\ref{gap}. This step is repeated by placing the ball in all possible empty areas, then repeated again with smaller balls. 

The original method uses a \emph{region growing} step to reach narrow corners, expanding the segments to adjacent pixels with similar color. This step relies on the different segments having different colors, which they do not in our monochrome input images. We therefore skipped this step, and instead expanded all segments to any locations reachable without crossing an edge line after completing the algorithm with ball size 2. Finally, we used a ball of size 1 to segment all remaining areas. Fig.~\ref{segmentation} shows a segmented image.

This change in the algorithm makes it more sensitive to the choice of initial ball size, as corners not reached by a larger ball may be filled by a smaller ball before the region growing step is reached. While this encourages smaller starting ball sizes, any gap wider than the starting ball diameter will leak, essentially putting a lower bound on the starting size. If the target image has a few wide gaps, it can be worthwhile to manually close them with a few quick lines. For some of the results shown in this paper, we have added up to four lines to the target image for this reason.

We leave the choice of initial ball size up to the user, recommending a value between 2-5 pixels.

\subsection{Colorization}

The remaining colorization and post-processing happens in 4 steps.

\subsubsection{Select color per segment}
We enlarge the cGAN output to the same dimensions as the target image. For each segment found in the previous step, the mean color of the corresponding pixels of the cGAN image in RGB space is selected as the color for the entire segment. Two other selection methods were considered: selecting the most prominent color in the segment, and selecting one of the most common colors from the cGAN output image. In the end however, selecting the mean color gave good results at a low level of complexity.

\subsubsection{Increase saturation}
The averaging in the previous step can sometimes render the colors slightly washed out compared to the training image, due to colors in the cGAN colorization output floating out between segments. To combat this, saturation can be increased in the HSV colorspace. The optimal amount varies by image, and can be left up to the user. We found that values in the range 5-10\% produced satisfactory results. This step also helps separate the colors in RGB space for the next step.

\subsubsection{Color quantization}
To remove any remaining noise, color quantization can be performed to reduce the number of unique colors in the image. In Fig.~\ref{quant}, for example, the color quantization step consolidates the different shades of the character's legs to the same color. \emph{k} clusters of colors are found using k-means clustering in RGB space, and all pixels are colorized with the center value of their cluster. The optimal number of colors \emph{k} can differ between different images, so we leave it up to the user. The optimal number tends to lie in the interval 5-12. In many cases, this step is not necessary.

\subsubsection{Shading}
Finally, if the input image has screentones, we can use these to apply shading to the image. To achieve this, we first apply a strong Gaussian blur to the input image, smoothing out the sharp screentones. We saw good results using a kernel radius two pixels bigger than the one used for screentone removal in Section~\ref{sec:screentone}. We then darken the appropriate areas of the colorized image using the following on each pixel in each of the RGB channels:
$$ C_{[i, j]} = C_{[i, j]} - \frac{255 - M_{[i, j]}}{3},  $$
where $C$ is the colorized image we are working on and $M$ is the blurred monochrome input image. 

\begin{figure}
\captionsetup{justification=centering}
\begin{subfigure}[t]{.3\linewidth}
    \includegraphics[width=\linewidth]{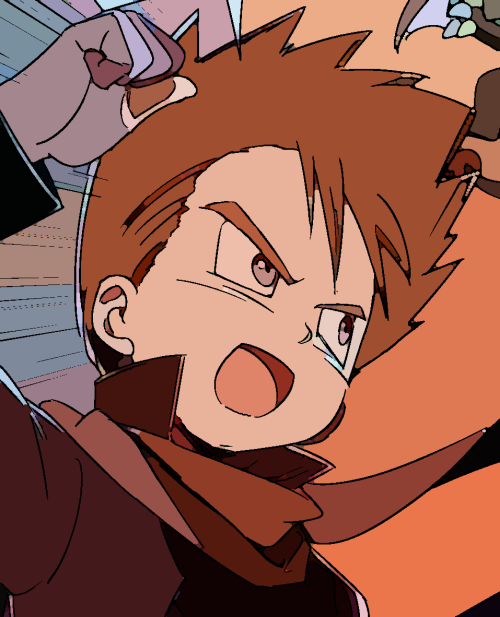}
    \caption{Target image had no screentones}
\end{subfigure}
\begin{subfigure}[t]{.3\linewidth}
    \includegraphics[width=\linewidth]{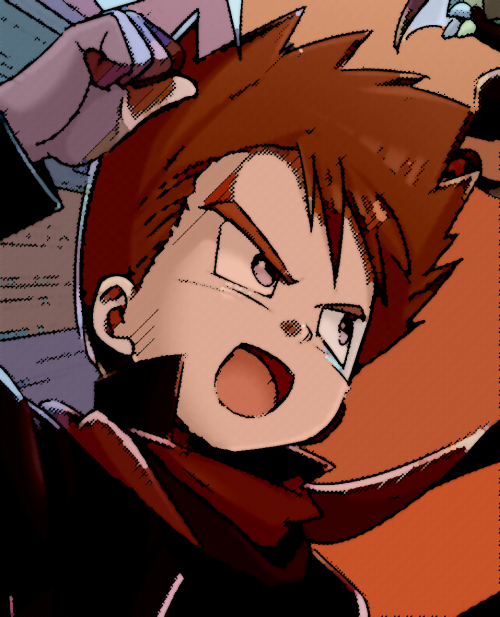}
    \caption{Target image had screentones}
\end{subfigure}
\begin{subfigure}[t]{.3\linewidth}
    \includegraphics[width=\linewidth]{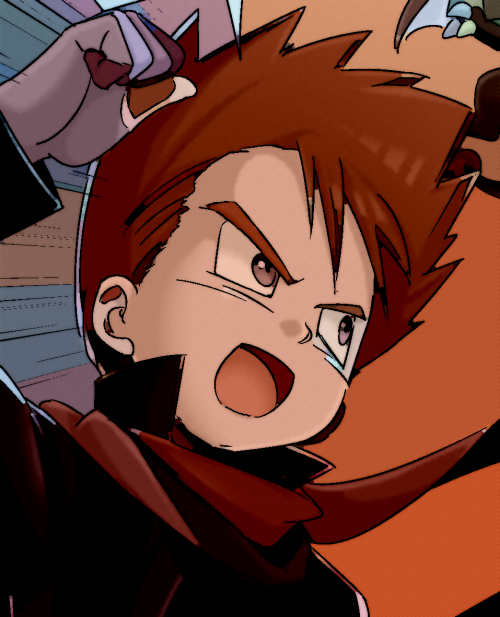}
    \caption{Both versions of target image supplied}
\end{subfigure}
\caption{Results using different versions of the target image. \label{toneorno}}
\end{figure}

\begin{table}
\centering
 \begin{tabular}{||c | c | c ||} 
 \hline
 Parameter & Permissible range & Recommended range \\ [0.5ex] 
 \hline\hline
 Gaussian blur radius & > 0 & 1 - 2\\\hline 
 Starting ball size & > 1 & 2 - 5 \\\hline
 Saturation increase & < 255 & 10 - 25 \\\hline
 Color clusters & > 0 & 5-20\\\hline
 \end{tabular}
 \caption{Tunable parameters and their ranges. \label{tunable}}
\end{table}

\section{Results}

\begin{figure}
\begin{subfigure}[t]{.3\linewidth}
\caption*{Training image}
\end{subfigure}\hfill
\begin{subfigure}[t]{.65\linewidth}
\caption*{Our results}
\end{subfigure}
\includegraphics[width=\linewidth]{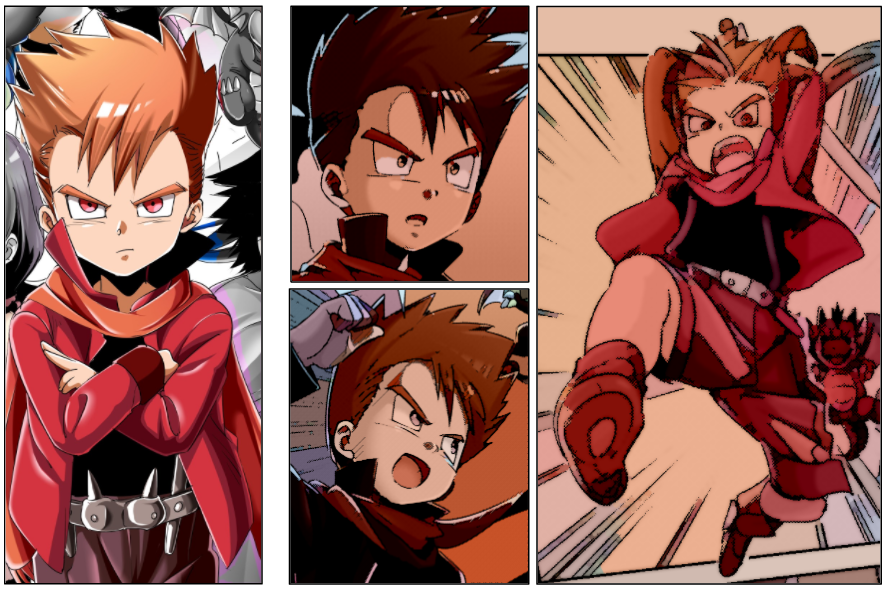}
\begin{subfigure}[t]{.30\linewidth}
\caption*{Training image}
\end{subfigure}\hfill
\begin{subfigure}[t]{.65\linewidth}
\caption*{Our results}
\end{subfigure}
\includegraphics[width=\linewidth]{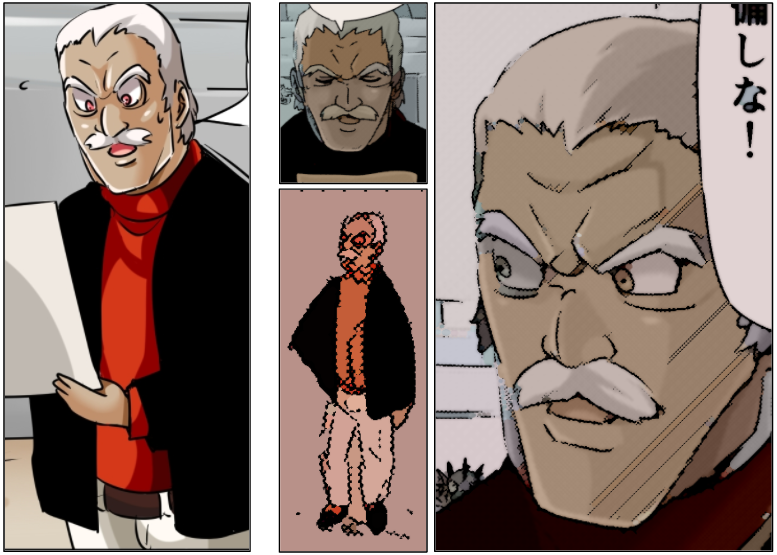}
\caption{Our results for Monster Beat dataset. On the left a colorized training image. On the right, images colorized by our method. \label{resG}}
\end{figure}

\begin{figure}
\begin{subfigure}{.45\linewidth}
\includegraphics[width=\linewidth]{mor.jpg}
\caption{Training image}
\end{subfigure}
\begin{subfigure}[t]{.45\linewidth}
\includegraphics[width=\linewidth]{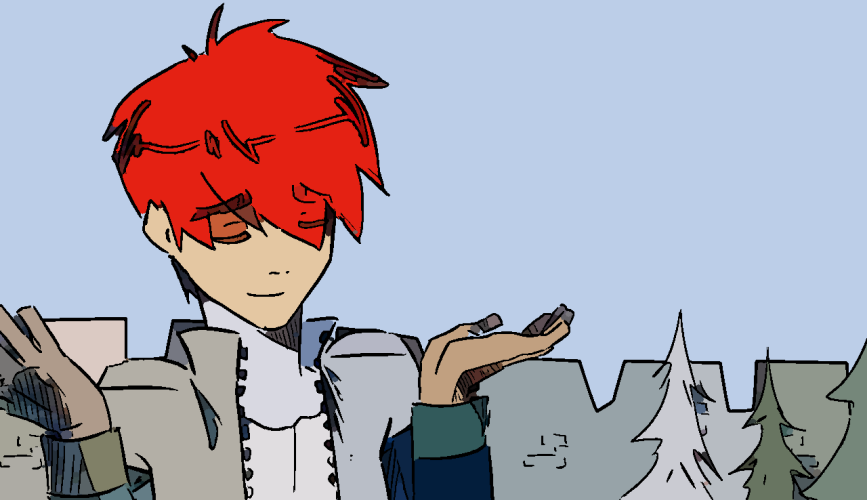}\\
\includegraphics[width=\linewidth]{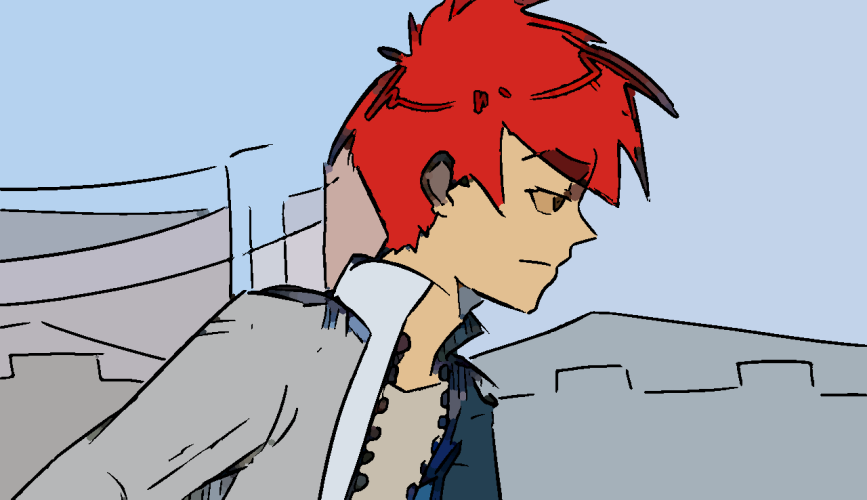}
\caption{Our results \label{morresultb}}
\end{subfigure}
\begin{subfigure}{.45\linewidth}
\includegraphics[width=\linewidth]{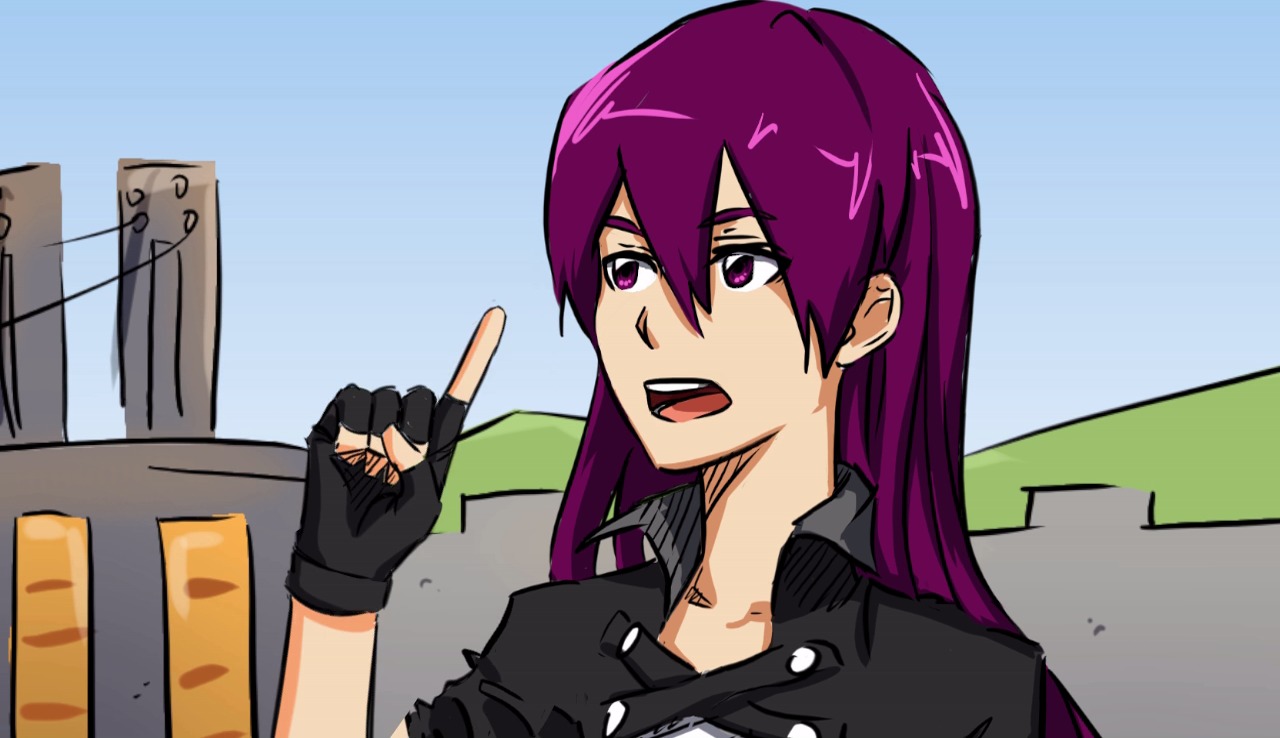}
\caption{Training image}
\end{subfigure}
\begin{subfigure}[t]{.45\linewidth}
\includegraphics[width=\linewidth]{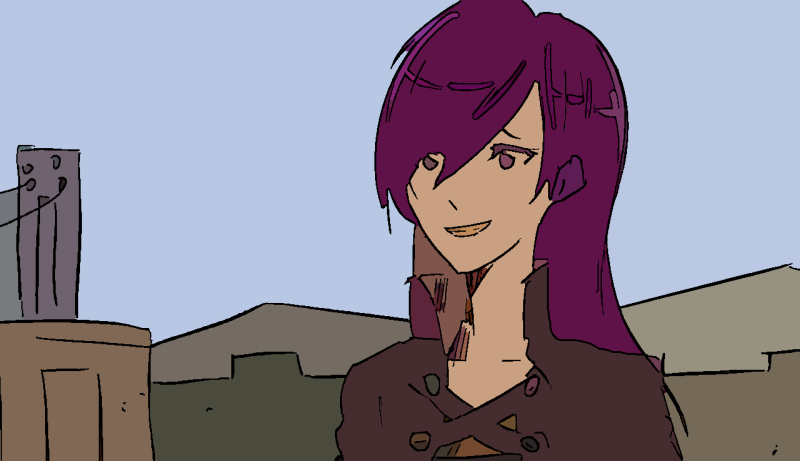}\\
\includegraphics[width=\linewidth]{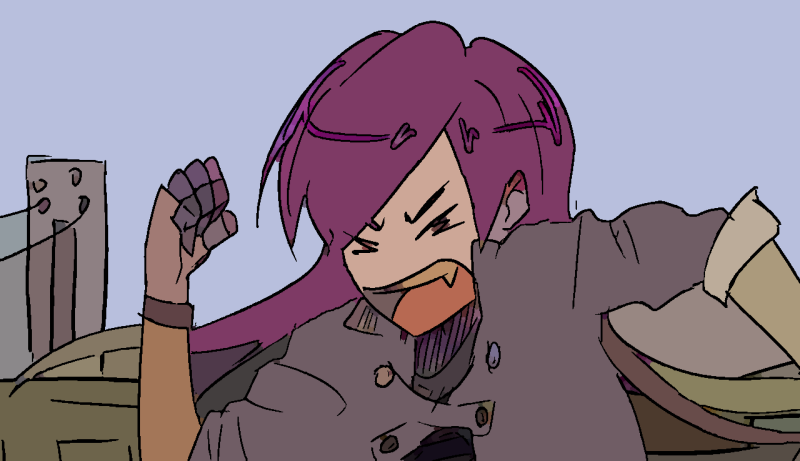}
\caption{Our results \label{morresultb2}}
\end{subfigure}
\caption{Our results for the Morevna dataset.  \label{morresult}}
\end{figure}

Our results are shown in Fig.~\ref{potential},~\ref{toneorno},~\ref{resG}, and~\ref{morresult}. 

In the cGAN colorization step, high-level structures such as hair, skin, and clothes are recognized and assigned the correct color in a translation-invariant manner. Remaining inaccuracies are mitigated by our subsequent post-processing methods. Comparing the cGAN output on the bottom left in Fig.~\ref{sty} to the output after post-processing in Fig.~\ref{morresultb} demonstrates the effectiveness of our segmentation technique. While the cGAN output's colors float out outside the edge lines, our post-processing confines the colors to their appropriate segments and produces a clean output. 

As we showed in Fig.~\ref{traindata}, training the cGAN on a cropped version of the training image matching the target image improves the results. If all colors expected in the target image are not present in the training image, the cGAN will not have enough information to colorize it correctly (upper triangle). On the other hand, if more colors are present in the training image than are expected in the target image, the cGAN tries to fit those colors in the image anyway, leading to unwanted colors in the result (lower triangle). 

When the target image has the same type of screentones as the training image, the colorization gets much more accurate, as it allows a mapping between screentone and color. In this way, the screentones help the colorization in much the same way as greyscale coloring would. When the target image does not have screentones, the quality of the results decrease. Example results from the Morevna dataset in Fig.~\ref{morresult} show that certain structures, such as hair, sky, and to some extent hands, were easily identified, but other parts were more problematic. For the first training image, we noted overfitting for a dark blue color in the lower middle of the image, but saw that the red hair was accurately colorized regardless of position in the image. 

In many of our results, the edge lines are slightly rough and extra lines have appeared around shaded areas. This is a side effect of the screentone removal step. In cases where the target image lacked screentones to begin with, such as in Fig.~\ref{morresult}, this phenomenon is not present. However, in those cases there can be no shading. If both versions of the target image is supplied, i.e. with and without screentones, the screentones can be used for shading while keeping the clean lines from the screentone-free version. A comparison is shown in Fig.~\ref{toneorno}.

\subsection{Segmentation}
\begin{figure}
\includegraphics[width=0.4\linewidth]{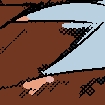}
\caption{Fail case for segmentation: leaking in the tip of the hair on top right, and extra segments due to remaining screentones in lower middle. \label{pointy}}
\end{figure}

The \emph{trapped ball} segmentation approach worked reasonably well. The resistance to small gaps in the edge lines proved useful when working with our dataset. It did however give rise to issues in narrow areas with gaps. After all areas that could fit a \emph{ball} with a diameter of 2 pixels were assigned, all existing segments got expanded to all non-assigned pixels reachable without crossing an edge line, referred to as the \emph{region expansion} step. This made larger segments float into areas that were too narrow to have been filled, but that had gaps in their outline. This was common in for example pointy strands of hair, as seen in Fig.~\ref{pointy}. Despite this, we opted to use the region growing step, as forgoing it gave rise to tiny segments being formed in all corners of the ground truth segments, resulting in a more mottled colorization.

The segmentation also had troubles in areas where the screentone removal was not fully effective, forming small dots of different colors in an area that should have been uniformly colored. An example is again in Fig.~\ref{pointy}.  

\subsection{Comparison with Style transfer}
\begin{figure}
\includegraphics[width=.35\linewidth, height=.35\linewidth]{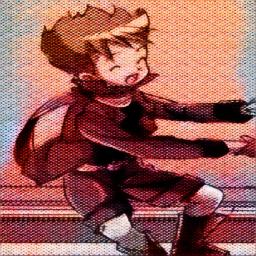}
\includegraphics[width=.35\linewidth, height=.35\linewidth]{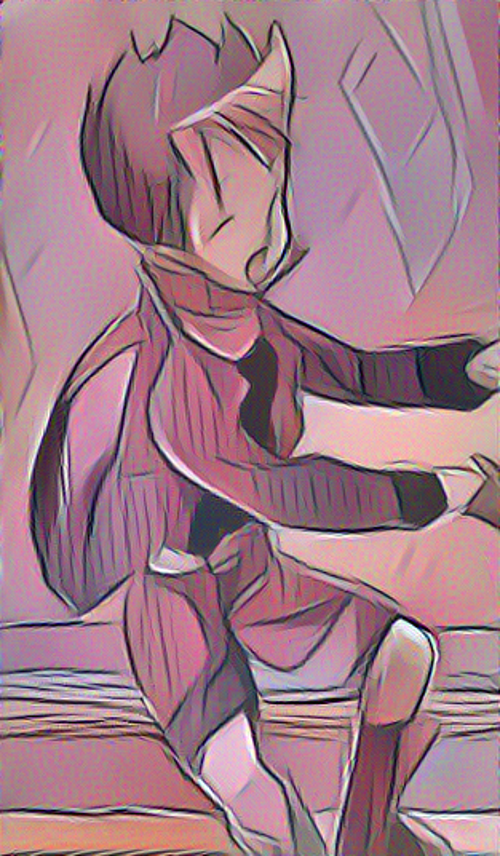}\\
\includegraphics[width=.35\linewidth, height=.35\linewidth]{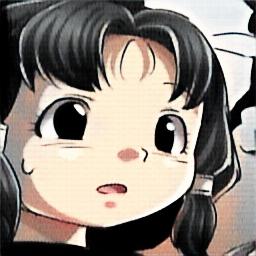}
\includegraphics[width=.35\linewidth, height=.35\linewidth]{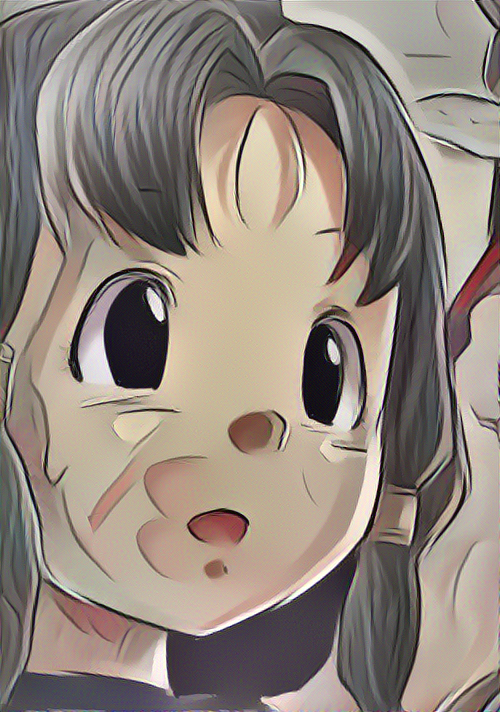}\\
\includegraphics[width=.35\linewidth, height=.35\linewidth]{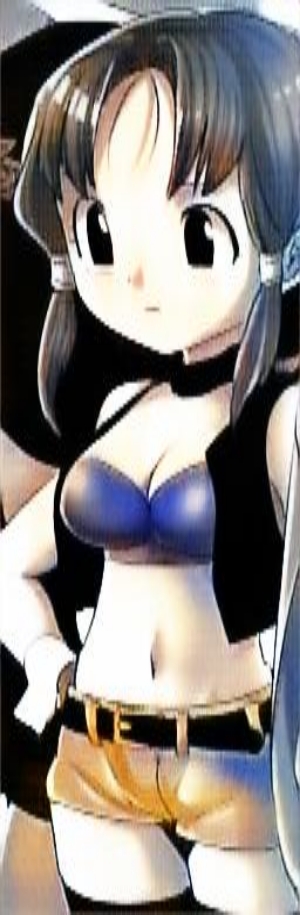}
\includegraphics[width=.35\linewidth, height=.35\linewidth]{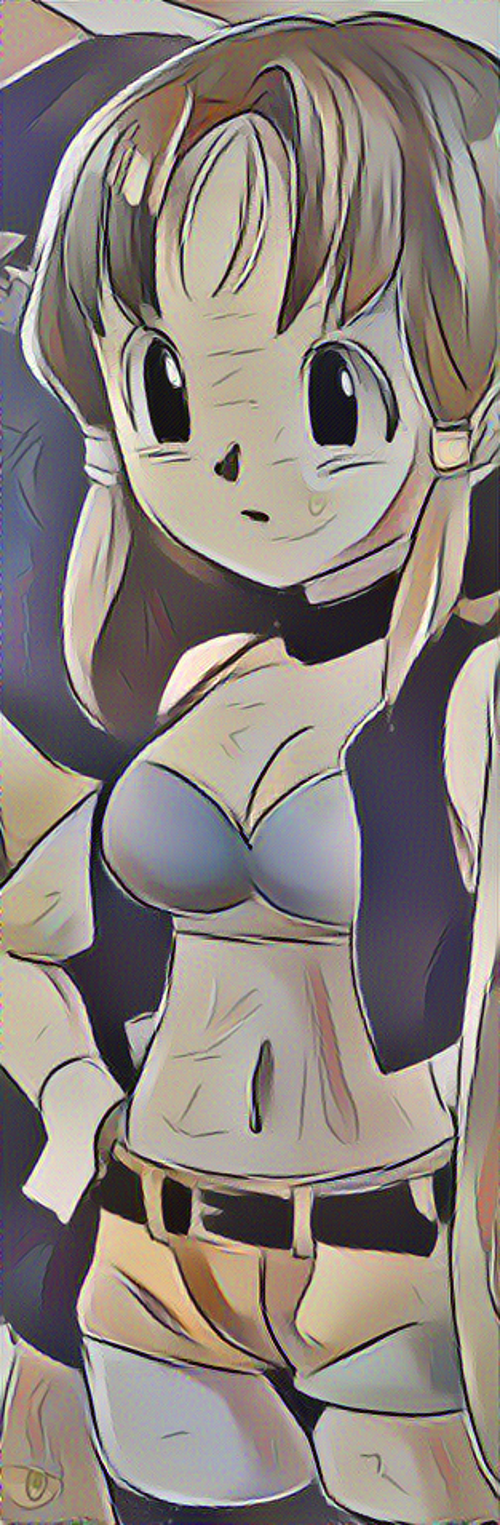}\\
\includegraphics[width=.35\linewidth, height=.35\linewidth]{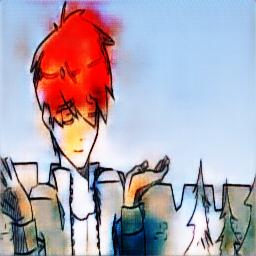}
\includegraphics[width=.35\linewidth, height=.35\linewidth]{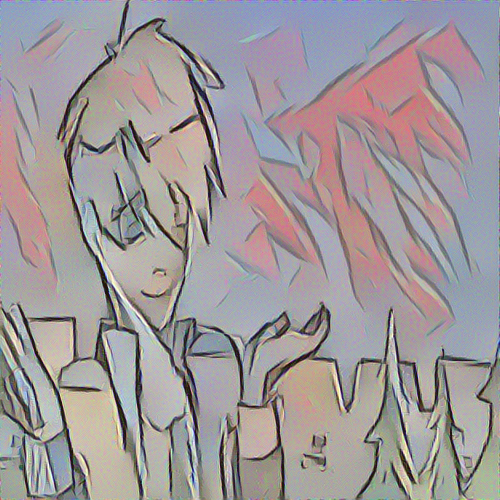}\\
\caption{Comparison between cGAN colorization (left) and style transfer (right) output. \label{sty}}
\end{figure}

We compare our cGAN colorization results to that of neural network-based style transfer. We use a combination of \cite{hertzmann2001image} and \cite{imgSynth} as implemented by Wentz\cite{imgana}, using the VGG16 features\cite{vgg16} for matching instead of looking at pixels directly. We keep the screentones in the images for this method. Results from this method are compared to our results in Fig.~\ref{sty}.

While the style transfer method is also mostly correct in color placement, especially with screentones present for guidance, the colors produced are pale. When no screentones are present, as in the image from the Morevna dataset on the bottom, our method was much more successful in locating important areas such as hair and skin. 

The greatest downside of the style transfer method is that each pair of images requires a long time to process, normally at least 5 minutes. To colorize \emph{n} images of a character's face, the style transfer method would need to be run \emph{n} times while our method would only need to be trained once. If the task is colorizing an entire manga volume, where a single character can easily appear over a hundred times, our method is much faster.

\section{Conclusion}
We have shown a method for colorizing manga images, requiring only a single colorized reference image for training. The results are clear, sharp and in high resolution, and stay true to the character's original color scheme.

Our method trains a cGAN on a single colorized reference image of a manga character, and uses the model to colorize new images of the same character. Blur, ambiguities and other problems in the output are mitigated by a segmentation and color-correction method. In many cases, the results of the cGAN colorization are good even as stand-alone results. 

While the current training time is very fast, it might be possible to get even faster results by using a cGAN pre-trained on a slightly larger dataset and finetuning it on the character-specific reference images. For a user wanting to colorize many different characters, this could save some extra time. This would be interesting to explore in future work.

For manga colorization, the biggest constraint is often the lack of available data, due to copyright restrictions. Methods using very low amounts of training data are thus very important in this field. By using training data featuring the same character as the target images, our method produced compelling results using only a single training image.

\begin{acks}
  The authors would like to thank Whomor for the use of their images.
\end{acks}


\end{document}